\documentstyle[11pt,rotate,psfig,paspconf,epsf]{article}

\def\solar{\ifmmode_{\mathord\odot}\else$_{\mathord\odot}$\fi~}
\def\deg{\ifmmode $\setbox0=\hbox{$^{\circ}$}$^{\,\circ}
          \else    \setbox0=\hbox{$^{\circ}$}$^{\,\circ}$\fi\,}

\begin{document}

\title{VLBI Observations of the Galactic Center Source 
Sgr\,A* at 86 GHz and 215 GHz}

\author{T.P. Krichbaum, A. Witzel, and J.A. Zensus}

\affil{Max-Planck-Institut f\"ur Radioastronomie, 
Auf dem H\"ugel 69, D-53121 Bonn, Germany}

\begin{abstract}
We{\footnote{for collaborators see acknowledgement}}
summarize previous VLBI observations of Sgr\,A* at
millimeter wavelengths and present new results from VLBI observations at 86\,GHz 
and at 215\,GHz. At 86\,GHz the measured
closure phase is close to zero, consistent with a point-like or symmetric
structure of $190 \pm 30$\,$\mu$as size. At 215\,GHz we have (for the
first time) detected Sgr\,A* with a signal-to-noise
ratio of 6. This yields a tentative size estimate of 
$50 \leq \theta \leq  190$\,$\mu$as, which is larger 
than the scattering size of $20 \mu$as at this frequency. 
The intrinsic size of Sgr\,A* thus appears to be a few up to a few ten
Schwarzschild radii (for a $2.6 \cdot 10^6$ M\solar black hole).
\end{abstract}


\section{Introduction}

At cm-wavelength the apparent size of the compact VLBI component of the
Galactic Center Radio Source Sgr\,A* decreases approximately quadratically
towards shorter wavelengths. This is usually interpreted as angular
broadening caused by diffractive scattering by the interstellar electrons.

The location at the dynamical center of the Galaxy, the compactness
and high brightness temperature, and the kinematics of the moving gas
and the stars around Sgr\,A* all strongly support the idea that 
Sgr\,A* is associated with a supermassive black hole with a
mass of $2.6 \cdot 10^6$ M\solar (eg.\  Genzel and Eckart, this conference).
At $D=8$\,kpc distance, the size of Sgr\,A* must at least be of order of
two Schwarzschild radii, which translates into $\theta \geq 2$\,R$_s/D = 13 \mu$as. 
This size becomes equal to the scattering size near $\lambda =1$\,mm.

Future mm-VLBI observations at $\lambda \leq 3$\,mm ($\nu \geq 86$\,GHz)
will therefore allow to image the direct environment of the black hole (and its event
horizon) with an angular resolution of only a few times the Schwarzschild radius
at frequencies, where interstellar scattering is not affecting the observed brightness
distribution.

The investigation of the compact radio source Sgr\,A* with mm-VLBI thus is mainly 
motivated by these two facts: high angular resolution and clear view.
In the following we summarize recent results obtained from VLBI observations
at 43\,GHz and 86\,GHz and report on the first VLBI-detection of Sgr\,A*
at 215\,GHz (single baseline). VLBI observations at these high frequencies are
difficult and at present 
are still affected by the limited number of available antennas, a sparsely
sampled uv-coverage and residual calibration uncertainties caused by 
atmospheric fluctuations (which increase towards shorter wavelengths),
and impurities of the receivers
(see also the remarks by Bower et al.\ and Doeleman et al.\ in this conference).
Therefore some of the results presented here still require
a careful interpretation and at least for the highest observing frequencies
need confirmation by better and more extended experiments.

\begin{table}[t]
\caption{Summary of size measurements from mm-VLBI.}

\vspace{0.5cm}
{\footnotesize
\begin{tabular}{lccccll}
$\lambda$& Axis &    Ratio    &$p.a.$     &  $S_{VLBI}$ &Epoch   & Reference  \\
$[cm]$  &$[mas]$&            &$[deg]$   & ~$[Jy]$  &$[~yrs]$   &               \\
      &     &            &       &          &        &                         \\ \hline
      &     &            &       &          &        &                         \\
0.69  &0.70$\pm$ 0.01&0.83$\pm$ 0.1&~87$\pm$ ~8 & 1.03$\pm$ 0.01&1997.10 &Lo et al. 1998           \\
0.69  &0.76$\pm$ 0.04&0.73$\pm$ 0.1&~77$\pm$ ~7 & 1.28$\pm$ 0.10&1994.75 &Bower et al. 1998        \\
0.69  &0.74$\pm$ 0.03&0.54$\pm$ 0.2&~90$\pm$ 10 & 2.10$\pm$ 0.10&1992.62 &Backer et al. 1993       \\
0.69  &0.70$\pm$ 0.10&0.60$\pm$ 0.2&115$\pm$ 20 & 1.42$\pm$ 0.10&1992.40 &Krichbaum et al. 1993$^a$\\
0.69  &0.75$\pm$ 0.08&1.0     &       & 1.42$\pm$ 0.10&1992.40 &Krichbaum et al. 1993$^b$\\
      &     &            &       &          &        &                         \\
0.35  &0.19$\pm$ 0.03&1.0     &       & 1.80$\pm$ 0.30&1995.18 &Krichbaum et al. 1997    \\      
0.35  &0.15$\pm$ 0.05&1.0     &       & 1.40$\pm$ 0.20&1994.25 &Rogers et al. 1994       \\       
0.35  &0.33$\pm$ 0.14&1.0     &       & 1.47$\pm$ 0.75&1993.27 &Krichbaum et al. 1994    \\ 
0.35  &0.22$\pm$ 0.19&1.0     &       & 1.25$\pm$ 0.35&1993.27 &Krichbaum et al. 1994$^c$\\
      &     &            &       &          &        &                         \\
0.14  &0.15$\pm$ 0.04&1.0     &       & 3.1$\pm$ 0.1~&1995.17 &Krichbaum et al. 1998$^d$\\
0.14  &0.11$\pm$ 0.06&1.0     &       & 2.0$\pm$ 0.2~&1995.17 &Krichbaum et al. 1998$^e$\\
\end{tabular}

\vspace{0.5cm}
Notes: \\
$^a$: size of the main component B1 in a 2 component model fit \\
$^b$: size from a single circular Gaussian component model fit \\
$^c$: reanalysis, using the assumption of NRAO\,530 being completely unresolved \\
$^d$: case 1, see text and Krichbaum et al. 1998 \\
$^e$: case 2, see text and Krichbaum et al. 1998 \\
}
\end{table}

\section{43\,GHz Observations}

In 1991--1992 the first VLBA antennas became operational at 43\,GHz. 
Subsequently the Galactic Center source Sgr\,A* was
detected for the first time with VLBI at this frequency (Krichbaum et al.\ 1993). 
The data revealed a compact core component B1 of 0.7\,mas size and some evidence
for an elongation of the overall source structure to the north, along p.a.$=-25$\deg.
With only 3 VLBA antennas (a 4$^{\rm th}$ antenna yielded only few detections) 
and one closure phase available, 
this experiment did not allow to unambiguously determine a possible elongation,
which seemed to be present nearly parallel to the main axis of the elliptical observing beam.
Follow-up 43\,GHz-experiments performed in 1992.6 by Backer et al.\ (1993) and in 1994.8
by Bower et al.\ (1998) with more VLBA antennas (N$=5$) revealed consistently a point source 
and a refined and more accurate size estimate. From these data,
the brightness distribution of Sgr\,A* is best 
described by a single and elliptically shaped Gaussian component 
(major axis: 0.75\,mas, axial ratio: $\sim 0.6$, cf. Tab.\ 1),
which is oriented in east-west direction. This orientation is also seen at the longer cm-wavelengths
(eg. Lo et al.\ 1985, Alberdi et al.\ 1993) and seems to be wavelength independent (Fig.\ 2). 
New multifrequency data from an experiment with 6 telescopes 
(5\,x\,VLBA + VLA$_1$) were presented during this conference
(Lo et al.\ 1998, and this proceedings). They reveal -- besides the aforementioned  east-west orientation --
a minor source axis, possibly larger than the extrapolated scattering size: 
at 43\,GHz the axial ratio seems to be $\sim 0.8$, 40-50\,\% greater 
than at lower frequencies. After subtracting the image broadening due to scattering, 
the authors interpret this as being caused by an intrinsic elongation of Sgr\,A*  
towards the north, not far off the orientation seen in the first 43\,GHz experiment in 1992.

In Table 1 we summarize the results of all 43\,GHz experiments on Sgr\,A*  published so far. The inspection
of the individual measurements reveals a very good agreement of the sizes within their  
uncertainties. We note an indication of flux density variability of the VLBI 
component, particularly if the data of 1992.6 and 1997.1 are taken into account.
We also note that already in the earliest 43\,GHz experiments evidence for an  east-west elongation 
of the core component was present (component B1 in Krichbaum et al., 1993). 
More detailed VLBI monitoring will be necessary to 
investigate whether the structure of Sgr\,A* is variable and if the flux density changes 
are real and related to this.

\section{86\,GHz Observations}

The 3\,mm (86\,GHz) VLBI observations of Sgr\,A* are even more difficult to perform and
to analyze than at 43\,GHz.
The stronger influence of the atmosphere (short coherence times of $10-20$\,sec) and the 
low aperture efficiency or small size  of some of the
antennas, still limit the detection sensitivity to $0.2 - 0.4$\,Jy ($7 \sigma$). Up to now, all of the
3\,mm VLBI observations of Sgr\,A* revealed visibility data on it only for 1--3
interferometer baselines. This restricts the determination of the brightness 
distribution to a circular component, precluding the determination of the axial
ratio of an eventually elliptical structure.

In 1993.27, Sgr\,A* was detected with VLBI at 86\,GHz on the $350$\,M$\lambda$ 
baseline between the 30\,m antenna
at Pico Veleta (Spain) and the 100\,m telescope at Effelsberg (Krichbaum et al.\ 1994). 
Although the signal-to-noise ratio of the detection was reasonably high (SNR $\simeq 20$), 
calibration uncertainties made the size estimate difficult. Depending of the details of
the calibration, we obtain a size of $0.33 \pm 0.14$\,mas, if the `a priori' calibration
and atmospheric opacity corrections are used, and a size of $0.22 \pm 0.19$\,mas with the
additional assumption that a secondary calibrator (NRAO\,530) is an unresolved point 
source\footnote{new VLBI images of NRAO\,530 show a more complex core-jet structure (Bower et 
al.\ 1997).} (see Table 1). 

In 1994.25 a VLBI experiment with three stations yielded data for Sgr\,A* on the
$250$\,M$\lambda$ baseline Kitt Peak to Owens Valley. This and the lack of a detection
on the longer baselines to the East Coast (Haystack Observatory) led to a more accurate
size measurement of $0.15 \pm 0.05$\,mas (Rogers et al.\ 1994).

With measurements of the visibility on at least three baselines, information on the
source orientation and possible deviations from circular symmetry can be obtained.
In March 1995 (1995.18) Sgr\,A* was observed on the triangle formed by the antennas
at Effelsberg, Pico Veleta
and a single antenna of the IRAM interferometer on Plateau
de Bure (Krichbaum et al.\ 1997 \& 1998). This yielded for the first time a direct measurement of the
closure phase, however only for a short time interval (5 scans, GST interval: 15h45 - 17h00).
The scan to scan variations of the closure phase are $\leq 15$\deg consistent with a
closure phase of zero in this hour angle range. 
This indicates a simple point-like or at least symmetric source
structure. The size derived from these observations ($0.19 \pm 0.03$\,mas) is consistent
with the size estimate of Rogers et al.\ (1994) and Doeleman et al. (this conference).
In Table 1 we summarize the size measurements at 86\,GHz.


\begin{figure}[t]
\plotfiddle{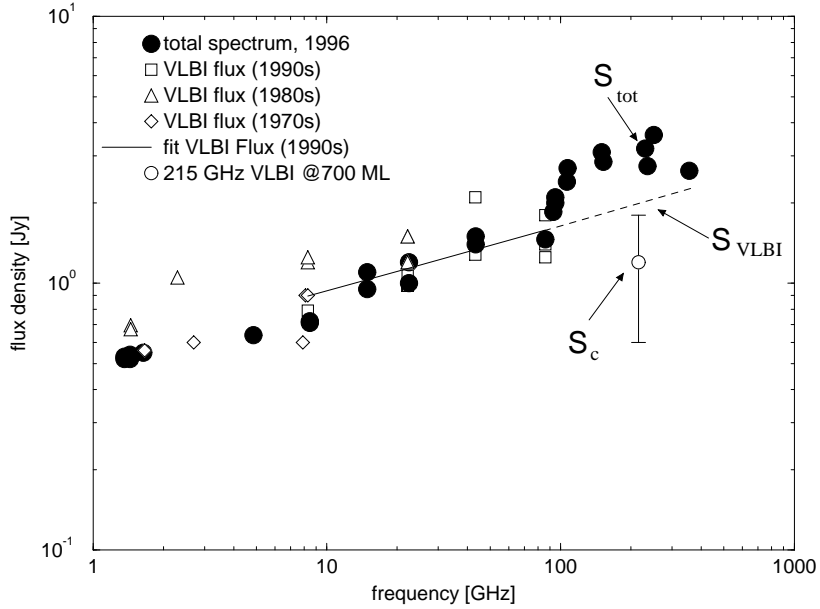}{8cm}{-90}{50}{50}{-210}{280}
\caption{\small
The cm- to mm-spectrum of Sgr\,A*. Filled circles show the total flux 
$S_{\rm tot}$. Open symbols denote fluxes of the VLBI component 
($S_{\rm{VLBI}}$) at 3 arbitrarily binned time intervals:
open squares for the 1990s, open triangles for the 1980s, open diamonds for the 1970s.
The solid line shows a power-law fit to the VLBI spectrum of the 1990s
($S_{\rm{VLBI}} \rm{[Jy]}= 0.535 \cdot \nu_{\rm GHz}^{+0.24}$), which compares well with
the total spectrum (filled circles). Above $\nu > 100$\,GHz the difference
between total flux and extrapolated spectrum (dashed line) indicates a flux density excess.
At 215\,GHz a correlated flux of $S_{c} = 1.2 \pm 0.6$\,Jy ($B= 700$\,M$\lambda$)
has been measured with VLBI (open circle).
}
\end{figure}

\section{215\,GHz Observations}

In 1994 it was demonstrated that successful VLBI observations at a wavelength of $\lambda=1.4$\,mm
are possible, using the two IRAM instruments at Pico Veleta and Plateau de Bure (Greve et al.\ 1995).
In a follow-up experiment performed in 1995, we clearly detected 6 out of 8 bright AGN
with scan averaged signal-to-noise ratios in the range of SNR$=7-35$ (Krichbaum et al.\ 1997).
In this experiment also Sgr\,A* was observed. A detailed analysis of the three VLBI 
scans (each of 6.5\,min duration) showed that Sgr\,A* was detected in 2
out of the 3 scans with a signal-to-noise ratio of SNR$=4.6$ and $6.0$, respectively
(Krichbaum et al.\ 1998).
With a formal detection threshold of SNR$=7$ for MK\,III VLBI data, this detection
at present should be regarded as marginal 
(for a detailed discussion see Krichbaum et al.\ 1998, and references therein). 
It is obvious that a confirmation by future 1\,mm-VLBI experiments is required. 

\begin{figure}[t]
\plotfiddle{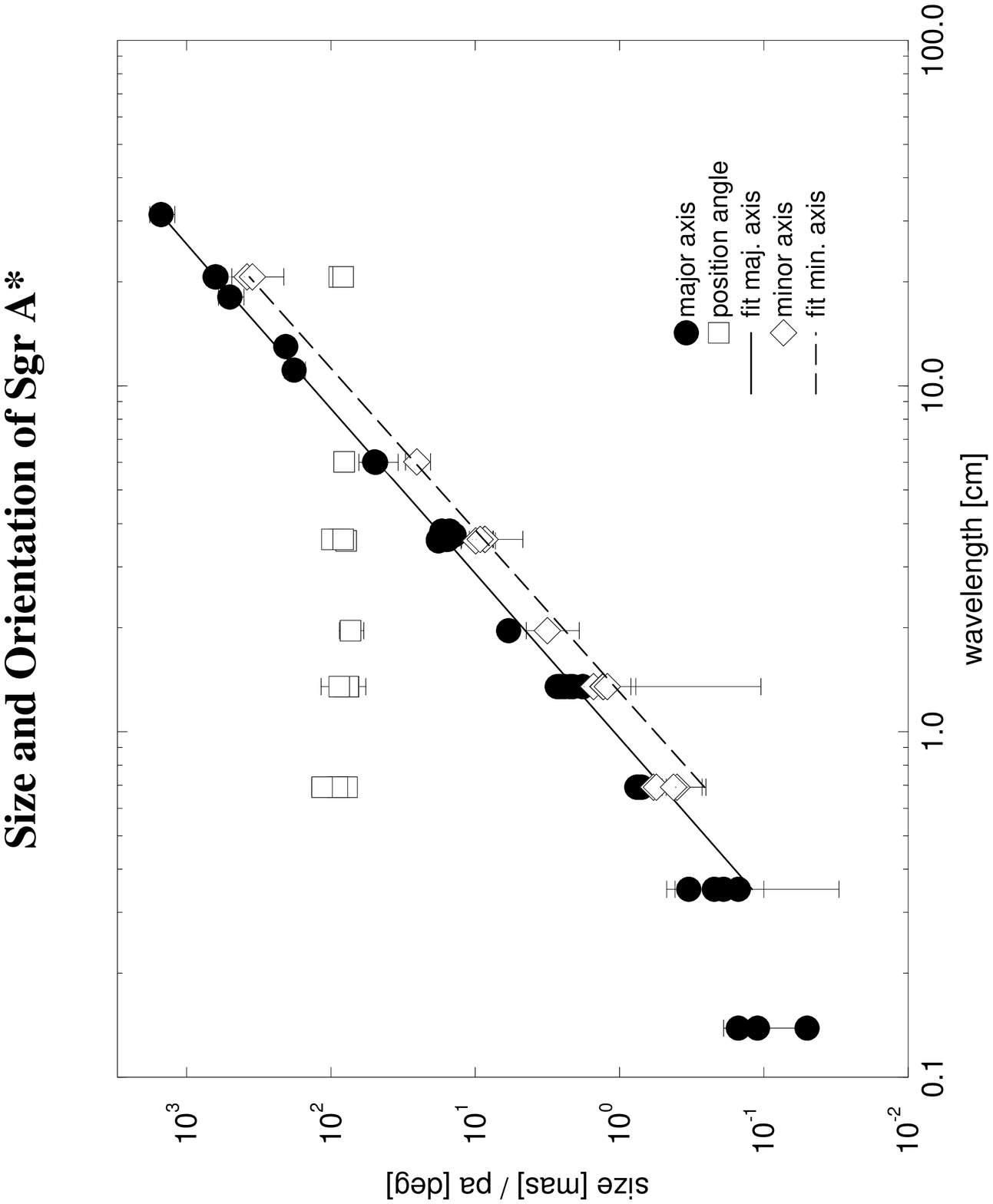}{8cm}{-90}{50}{50}{-210}{280}
\caption{\small
The major source axis (filled circles) of the
Gaussian VLBI component of Sgr\,A*, the minor source axis (open diamonds)
and the position angle of the major axis (open squares)
plotted versus wavelength. For the least squares fit to the major
axis only data for $\lambda \geq 3$\,mm were used (solid line), for the fit
to the  minor axis only data for $\lambda \geq 7$\,mm were
taken into account (dashed line).}
\end{figure}

If the detection is real, the correlated flux density for Sgr\,A* falls in the 
range of $S_c = 0.6 - 0.8$\,Jy if the `a priori' calibration is used and 
$S_c = 1.2 \pm 0.6$\,Jy, taking into account also residual systematic calibration
errors. Thus the correlated flux appears to be lower than the zero spacing flux extrapolated 
from VLBI measurements at longer wavelengths
($S_{\rm VLBI}=2.0 \pm 0.2$\,Jy, see solid line in Fig.\ 1), and also lower than the total flux 
$S_{\rm tot} = 3.1 \pm 0.1$\,Jy determined from flux density measurements in 1996 
(Falcke et al.\ 1998, filled circles in Fig.\ 1). We note that $S_{\rm tot}$ 
results from observations with observing beams of one to a few arcseconds, four orders of 
magnitude larger than the resolution achieved with VLBI at 215\,GHz. It is therefore 
possible that on intermediate angular scales structure components might exist, which are
presently not seen. We note that below 100\,GHz, $S_{\rm tot}$ and $S_{\rm VLBI}$ are
equal, excluding extended intermediate scale structures at longer wavelengths. 

\begin{figure}[t]
\plotfiddle{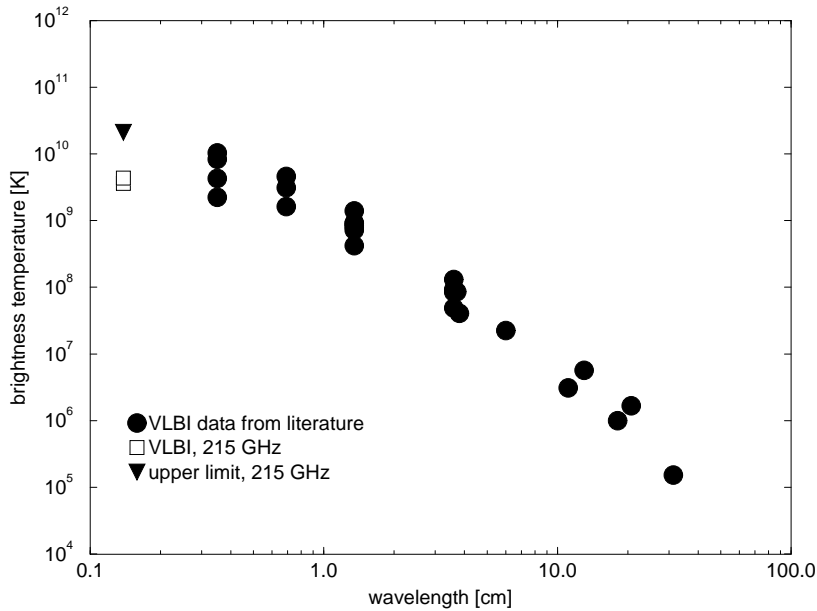}{8cm}{-90}{50}{50}{-210}{280}
\caption{\small
The apparent brightness temperature of the VLBI component of Sgr\,A* plotted versus wavelength.
Circles denote measurements at $\lambda \geq 3$\,mm. The filled triangle
denotes an upper limit, adopting $\theta \geq 50$\,$\mu$as
at 215\,GHz. The open squares represent the best estimate at 215\,GHz for 
`case 1' and `case 2' (see text and Table 1).}
\end{figure}

If the correlated flux of Sgr\,A* indeed is lower than the total flux,
the source would be resolved and a size could be estimated. As mentioned above,
above 100\,GHz there is a discrepancy between the total flux resulting from 
direct measurements with single antennas or local interferometers ($S_{\rm tot}$, see Fig.\ 1) and the flux density 
extrapolation resulting from VLBI observations ($S_{\rm VLBI}$, see Fig.\ 1). 
Using as total flux $S_{\rm tot}$, one 
obtains a size of $\theta_1 = 150 \pm 40$\,$\mu$as (case 1). If instead for the
total flux $S_{\rm VLBI}$ is used, a size of  
$\theta_2 = 110 \pm 60$\,$\mu$as is derived (case 2). Both sizes are close to the source size
seen at 86\,GHz (s. Table 1). In summary it is therefore likely that at 215\,GHz 
the size of Sgr\,A* lies in the range of $50 \leq \theta \leq  190$\,$\mu$as.

\begin{figure}[t]
\plotfiddle{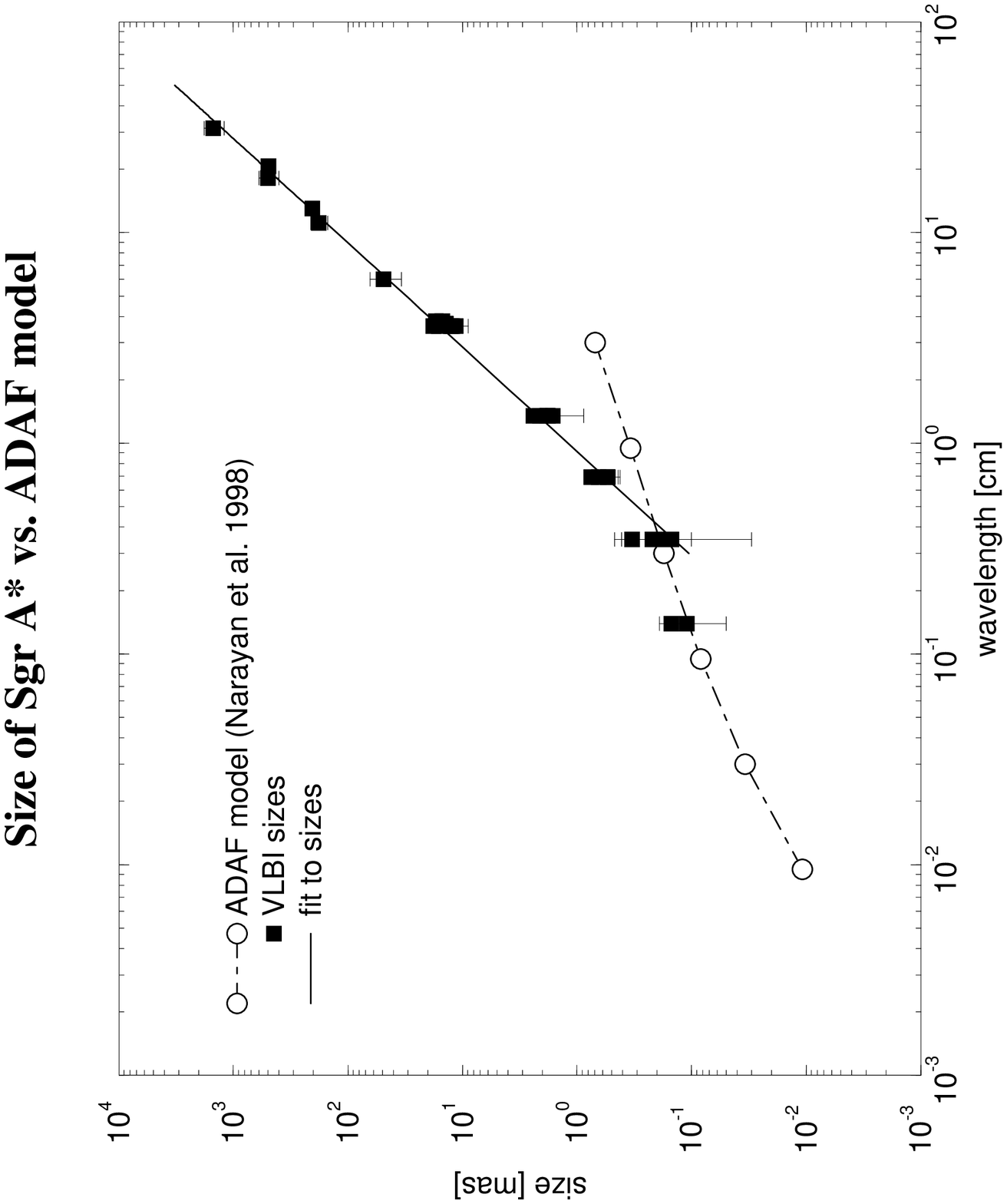}{8cm}{-90}{50}{50}{-210}{280}
\caption{\small
The geometric mean of major and minor axis of the VLBI component of
Sgr\,A* plotted versus wavelength.
Open symbols show the size predictions from the advection-dominated accretion flow
(ADAF) model of Narayan et al.\ (1998). At 1.4\,mm and 3\,mm the predicted and
observed sizes agree remarkably well.}
\end{figure}

\section{Discussion}

In Figure 2 the decrease of source size towards shorter wavelengths
is plotted, including all published VLBI measurements of the size
of Sgr\,A* and the new measurements presented in this paper.
It is obvious that the major and minor source axis obey almost 
the same wavelength dependence: fitting a power law 
($\theta \propto \lambda^n$) to the data, we obtain for the power law index 
a value of $n= 2.04 \pm 0.01$ for the major source
axis and $n= 1.96 \pm 0.04$ for the minor axis. Down to a wavelength of 1.3\,cm,
the orientation of the major axis seems to be independent of wavelength,
being close to 80\deg (open squares in Fig.\ 2). 
For shorter wavelengths, no data for the position angle
are yet available.

From the fit to the sizes, the scattering size at millimeter wavelengths can be
estimated empirically: for the major axis we obtain 0.5\,mas at 43\,GHz, 
0.12\,mas at 86\,GHz and 0.02\,mas at 215\,GHz.
These numbers do not change significantly, if instead of the major axis
the geometric mean of the minor and major axis is used (Fig.\ 4). The
comparison with the sizes given in Table 1 clearly shows that, starting from 
43\,GHz, the discrepancy between measured source size and expected
scattering size increases with frequency. It is therefore
likely that the scattering screen becomes more transparent towards shorter
wavelengths and that we are close to see the underlying intrinsic source 
structure of Sgr\,A*.
Further evidence for this comes from (i) the recently observed
increase of the axial ratio, which at 43\,GHz is probably larger
than at 22\,GHz (see Lo et al.\ 1998, and this conference), and
(ii) from an increase of the flux density variability amplitudes towards higher frequencies
(Zhao et al.\ 1991, Tsuboi et al.\, this conference).
Particularly the latter cannot be explained by the scattering effect.

Despite remaining measurement uncertainties at  215\,GHz,
the size of Sgr\,A* is at least a factor of $2-3$
larger than the scattering size ($20$\,$\mu$as) at this frequency. Further
evidence for the correctness of this size measurement comes from the
the brightness temperature $T_B$. In Figure 3 we
plot $T_B$ versus wavelength (for the calculation of $T_B$ 
we used all published size and flux density measurements from VLBI). 
If we extrapolate the brightness temperature from the longer wavelengths
to $\lambda = 1.4$\,mm (215\,GHz), we obtain $T_B \leq 4.5 \cdot 10^{10}$\,K. 
With a total flux of Sgr\,A* in the range  $S_{\rm tot} = 2-3$\,Jy and using $T_B$
from above,
this yields a size of $\theta \geq 30 - 40$\,$\mu$as, again larger
than the scattering size. 

A lower limit to the size of Sgr\,A* is given by the Schwarzschild
diameter ($2 R_s/D = 13$\,$\mu$as) of the central black hole. The observable
size of Sgr\,A*, however, should be much larger, if the radiating volume 
is related to the accretion disk, which for the inner stable orbits
should have a diameter of at least a few up to a few ten $R_s$. 
The broad band spectral properties and the low (sub-Eddington)
luminosity of Sgr\,A* are well explained by 
the advection-dominated accretion flow (ADAF) model of Narayan et al.\ (1998). 
This model predicts a size of $0.1$\,mas ($\sim 15 R_s$) at 215\,GHz.
In Figure 4 we display the sizes from the ADAF model together with 
the measurements. At 3\,mm and 1.4\,mm wavelength, the agreement is striking.

We note that the existing models predict slightly different variations of 
the intrinsic source size with the wavelength (eg. Melia et al., 1992, 
Falcke \& Biermann, 1998). This can be used to descriminate between
the models, as soon as a deviation of the size from the $\lambda^2$-law 
(scattering) will be confirmed.

Future 1\,mm VLBI experiments will be extremely important to
improve the accuracy of the size measurement and will help to
decide whether Sgr\,A* harbors a supermassive black hole or not.
A break-through probably can be achieved, if Sgr\,A* can be imaged
with 1\,mm-VLBI also on the longer (continental, or even transatlantic) baselines.
With an angular resolution of up to $20-30$\,$\mu$as reached in this
configuration, relativistic effects
near the event horizon of a supermassive black hole could be {\it directly}
studied. To provide the necessary sensitivity for this, the existing
(eg.\ BIMA, OVRO, Plateau de Bure) and future (eg.\ LSA, MMA) interferometers
need to participate as phased arrays in such VLBI experiments.

\acknowledgments
The results summarized here are obtained in a collaboration
between various observatories including in addition to the authors
the following scientists: 
D. Graham (MPIfR), A. Greve, M. Grewing, J. Wink (IRAM), F. Colomer, P. de Vicente, 
J. G\'omez--Gonz\'alez (Yebes), and A. Baudry (Bordeaux). 
We also like to appreciate very much the support of P.G. Mezger.

\end{document}